\documentclass[10pt,twocolumn,floatfix,preprintnumbers,superscriptaddress,nofootinbib,aps,prd]{revtex4-2}

\usepackage{graphicx}
\usepackage{amsmath,amssymb}
\usepackage{booktabs}
\usepackage{array}
\usepackage[utf8]{inputenc}
\usepackage[colorlinks,linkcolor=blue,citecolor=blue,urlcolor=blue]{hyperref}
\usepackage{url}
\usepackage{orcidlink}

\begin{document}

\title{Machine Learning for Invisible Dark Boson Searches at the Electron-Ion Collider}

\author{Rojae Mighty\,\orcidlink{0009-0002-5701-318X}}
\email{rojae.mighty@stonybrook.edu}
\affiliation{Department of Physics and Astronomy, Stony Brook University, Stony Brook, NY 11794-3800, USA}

\author{Ankush Reddy Kanuganti\,\orcidlink{0000-0002-0789-1200}}
\email{akanugant@bnl.gov}
\affiliation{Physics Department, Brookhaven National Laboratory, Upton, NY 11973-5000, USA}

\begin{abstract}
    We investigate whether adding $t$, the positive magnitude of the squared nuclear four-momentum transfer, enables boosted decision trees (BDTs) to improve invisible-dark-boson selection relative to optimized rectangular cuts at the Electron-Ion Collider. We model coherent exclusive scalar and vector production at generator level in electron--gold collisions at 18 GeV by 100 GeV per nucleon. Both methods use identical weighted samples, inputs, preselection, and optimization objectives. Using only electron information, the BDT provided no consistent advantage over optimized cuts for signal selection across 11 masses for each boson type. When both methods also use $t$, the BDT distinguishes signal from background slightly better than optimized cuts at 10 GeV for both boson types. These results motivate further investigation of machine learning in EIC dark-boson searches through exclusive processes where $t$ can be reconstructed.
\end{abstract}

\keywords{Electron-Ion Collider; invisible dark bosons; nuclear recoil; momentum transfer; boosted decision trees; event selection}

\maketitle

\section{Introduction}\label{sec:1}

The Electron-Ion Collider (EIC) at Brookhaven National Laboratory is designed to investigate the quark and gluon structure of protons and nuclei, including the origin of nucleon spin and the spatial and momentum distributions of their constituents \citep{ref7}. Its electron and ion beams also offer opportunities to search for light, weakly coupled particles beyond the Standard Model, complementing the central hadron-structure program \citep{ref1,ref2}.

Davoudiasl and Liu proposed a search for invisibly decaying scalar and vector dark bosons through coherent exclusive production, $e\,\mathrm{Au}\to e\,\mathrm{Au}\,\phi$, with the gold nucleus remaining intact \citep{ref1,ref2}. The scenario requires a weak but non-negligible electron coupling and a substantial invisible branching fraction. Since the invisible final state must be inferred from visible recoil, the original search applies mass-dependent rectangular cuts to four electron quantities: momentum transfer $Q^2$, transverse momentum $p_{T,e}$, pseudorapidity $\eta_e$, and energy $E_e$.

The nuclear momentum transfer enters the production calculation but is not used as an additional selection variable in the original search. We denote its positive squared magnitude by $t$. If the outgoing ion can be reconstructed, $t$ provides recoil information not fully determined by the electron when $\phi$ is invisible. Establishing whether this information improves selection can help prioritize studies of outgoing-ion acceptance and recoil resolution.

We compare boosted decision trees (BDTs) with a strong multistart rectangular selector. Reoptimizing the cuts on the same samples isolates the benefit of the BDT\textquotesingle s decision boundary from differences in the baseline selection. The electron-only comparison tests whether the BDT outperforms optimized rectangular cuts using recoil-electron information alone; adding $t$ tests whether it changes their relative performance.

We find practical equivalence between electron-only BDTs and optimized rectangular cuts, while adding $t$ produces a small, reproducible BDT advantage at 10 GeV for both signal spins. This result motivates studies of whether outgoing-ion reconstruction preserves the information responsible for the gain.

\section{Signal Model and Event Generation}\label{sec:2}

\subsection{Coherent exclusive production process}\label{sec:2.1}

We consider scalar and vector bosons coupled to electrons with benchmark coupling $g_{\phi e}$ = $10^{-4}$ and an invisible branching fraction treated as unity. The signal process is $e(k)+\mathrm{Au}(P)\to e(k')+\mathrm{Au}(P')+\phi(p_\phi)$, where the outgoing gold ion remains intact. The incoming electron energy is 18 GeV and the incoming gold energy is 100 GeV per nucleon. Gold is modeled with $A=197$ and $Z=79$, and a Helm elastic form factor describes the coherent nuclear response.

\begin{figure*}[!htbp]
\centering
\includegraphics[width=\textwidth,height=0.42\textheight,keepaspectratio]{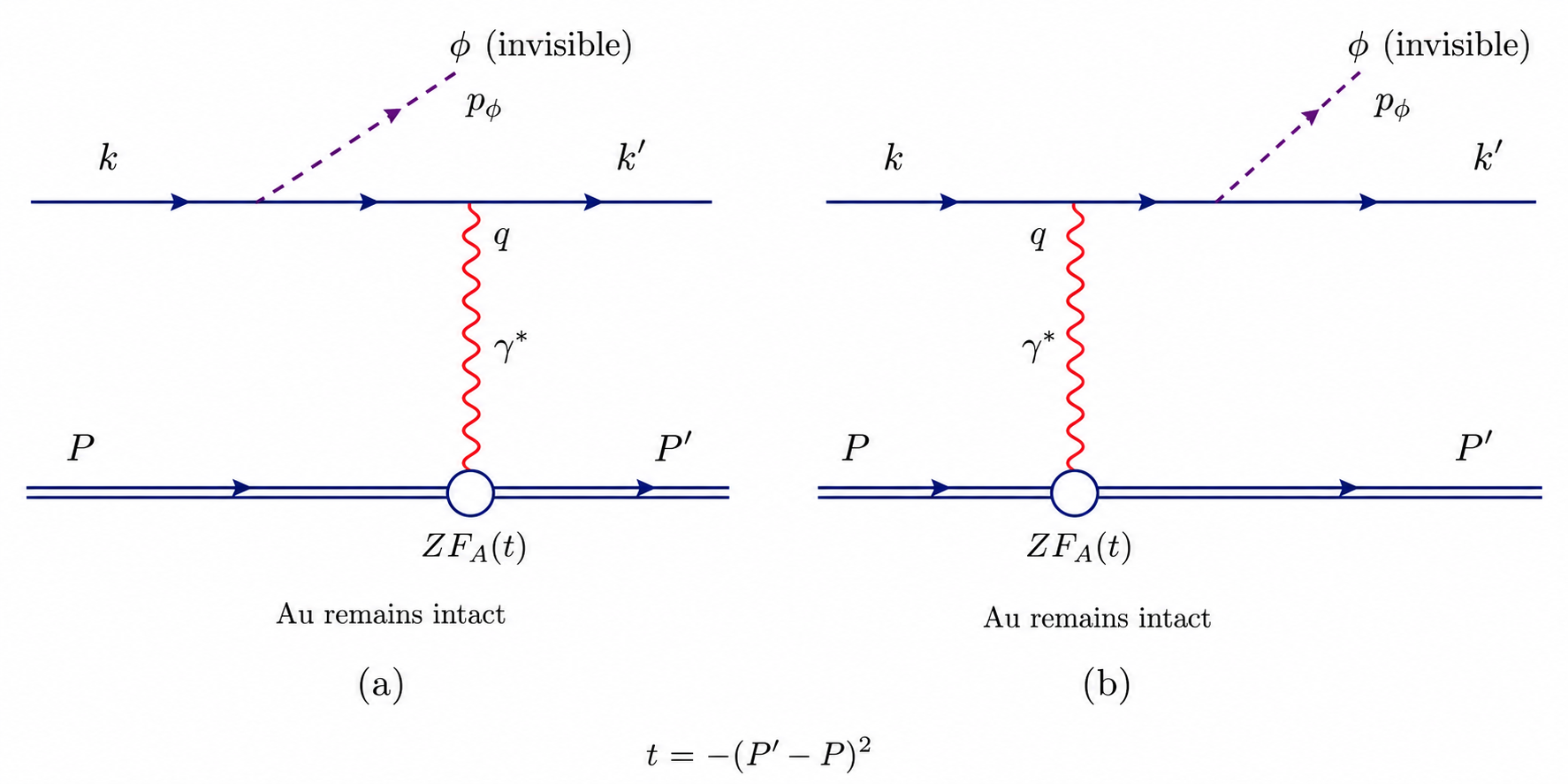}
\caption{Coherent exclusive $e(k)+\mathrm{Au}(P)\to e(k')+\mathrm{Au}(P')+\phi(p_\phi)$ production. The dark boson is emitted before panel (a) or after panel (b) virtual-photon exchange with intact Au. The nuclear vertex is $Z F_A(t)$, with $t=-(P'-P)^2$, and the amplitude coherently sums both diagrams. The dashed line denotes a scalar or vector $\phi$.}\label{fig:1}
\end{figure*}

Figure~\ref{fig:1} shows the two electron-line emission diagrams, whose amplitudes are summed coherently in the signal calculation.

\subsection{Kinematic definitions}\label{sec:2.2}

All cuts and classifier inputs are evaluated in the collider laboratory frame. We choose the incoming gold beam to define $+z$ and the incoming electron beam to travel in the opposite direction. Let $q_e=k-k'$ be the four-momentum lost by the electron. The positive electron momentum-transfer variable is $Q^2=-q_e^2$. The electron observables used by the reference selection are $Q^2$, $p_{T,e}$, $\eta_e$, and $E_e$.

The nuclear momentum transfer is defined using the positive spacelike convention, $t=-(P'-P)^2>0$.

Although $t$ already enters the production calculation, here it is also supplied to both selectors as an additional input. We use its exact generator-level value; experimental reconstruction would introduce acceptance, efficiency, and resolution effects.

\subsection{Event generation, weights, and cut efficiencies}\label{sec:2.3}

Signal events are generated by direct numerical integration of the exact three-body phase space and the cited scalar and vector matrix elements \citep{ref1,ref3,ref4}. The calculation includes the Helm form factor, the two electron-line emission amplitudes, their interference, and the full vector-polarization sum. Generated four-momenta are transformed to the collider laboratory frame before $Q^2$, $p_{T,e}$, $\eta_e$, $E_e$, $t$, and the electron-boson invariant mass are calculated.

Our TFoam implementation uses adaptive importance sampling to generate events with Monte Carlo weights $w_i$. Cross sections and efficiencies are calculated from sums of weights. For a selection $C$, the efficiency is $\epsilon_C=\sum_i w_i I_i(C)/\sum_i w_i$, where $I_i(C)$ equals one if event $i$ passes and zero otherwise.

The inclusive cross sections and those selected by the mass-dependent electron cuts in Table I of Ref.~\citep{ref1} are obtained from the same matrix element and phase-space measure, without an independent normalization fitted to the selected rates. Samples for the ML comparison are generated conditional on this preselection, and both the BDT and optimized rectangular selector act as second-stage refinements.

\subsection{Vector and scalar signal cross sections}\label{sec:2.4}

\begin{figure*}[!htbp]
\centering
\includegraphics[width=\textwidth,height=0.42\textheight,keepaspectratio]{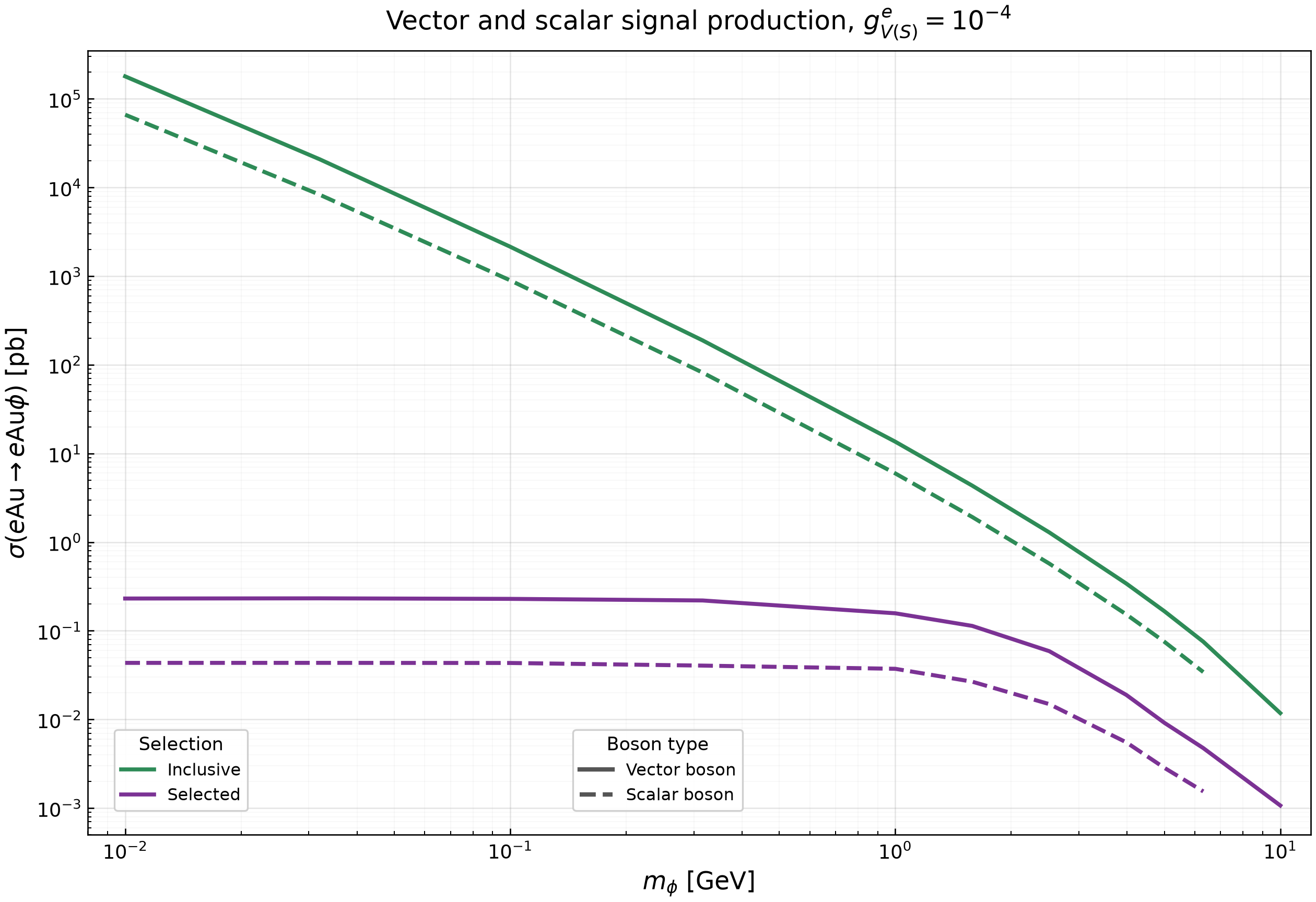}
\caption{Coherent vector and scalar signal cross sections as functions of boson mass for $g_{\phi e}$ = $10^{-4}$ at $18~\mathrm{GeV}\times100~\mathrm{GeV}$ per nucleon. Green and purple show our inclusive and selected calculations, respectively. Solid lines denote vector bosons and dashed lines denote scalar bosons. The selected curves apply the mass-dependent electron preselection described in Section~\ref{sec:2.3}. Both axes are logarithmic.}\label{fig:2}
\end{figure*}

Figure~\ref{fig:2} shows the inclusive and preselected scalar and vector production cross sections over the mass range 0.01--10 GeV for the setup in Section~\ref{sec:2.1}.

The inclusive cross sections decrease rapidly with boson mass, with the vector rate exceeding the scalar rate throughout the scan. Table~\ref{tab:1} gives the rates at $m_\phi=1$ GeV, where the preselection efficiencies are approximately 1.16\% for the vector and 0.625\% for the scalar. The selected samples provide the baseline for the subsequent BDT and rectangular-cut comparison.

\begin{table}[!htbp]
\caption{Inclusive and selected signal cross sections from our weighted numerical integration at $m_\phi$ = 1 GeV and $g_{\phi e}$ = $10^{-4}$. The selected rates include the mass-dependent electron preselection.}\label{tab:1}
\centering\small\setlength{\tabcolsep}{3pt}
\begin{tabular}{lrr}
\toprule
Signal & Inclusive [pb] & Selected [pb] \\
\midrule
Vector & 13.5868 & 0.157512 \\
Scalar & 5.96175 & 0.0372382 \\
\bottomrule\end{tabular}
\end{table}

\section{Background Model}\label{sec:3}

\subsection{Backgrounds treated in the reference analysis}\label{sec:3.1}

The reference paper identifies coherent photon bremsstrahlung, $e\,\mathrm{Au}\to e\,\mathrm{Au}\,\gamma$, as the dominant reducible background when the emitted photon is not detected. Its signal-like topology contains a recoil electron, an intact ion, and apparent missing energy. The paper assigns a miss probability of $10^{-6}$ to central photons and assumes that photons outside $|\eta_\gamma|$ = 3.5 are missed. These probabilities describe detector failure parametrically; they are not obtained from a full detector simulation.

The reference also evaluates deep-inelastic scattering, $eA\to eXj$, at parton level with MadGraph5\_aMC@NLO and assumes that a zero-degree calorimeter veto rejects 95\% of the relevant nuclear-breakup background. Its Table I lists effective photon and DIS cross sections after each mass-dependent electron selection. An irreducible neutrino-pair background is reported to be negligible \citep{ref1}.

\subsection{Coherent-photon proxy used in this study}\label{sec:3.2}

The public ancillary package provides one-dimensional photon distributions but no event-level samples for training. We generate our own photon events under the same beam kinematics, using an 18 GeV electron beam and a 100 GeV-per-nucleon gold beam.

We therefore construct a reproducible coherent-photon proxy from the massless-vector limit of our coherent matrix element, using electromagnetic coupling e = 0.3028221209 and a photon-mass regulator of $10^{-6}$ GeV. The regulator is far below every analyzed kinematic scale and only stabilizes the numerical representation of the massless limit. For each dark-boson mass hypothesis, the photon sample is generated with the same Table I electron preselection used for the corresponding signal comparison.

Each photon event carries its Monte Carlo integration weight multiplied by the photon-miss probability specified in Section~\ref{sec:3.1}. The generated distributions supply the background shapes used in the classifier comparison.

\begin{figure*}[!htbp]
\centering
\includegraphics[width=\textwidth,height=0.42\textheight,keepaspectratio]{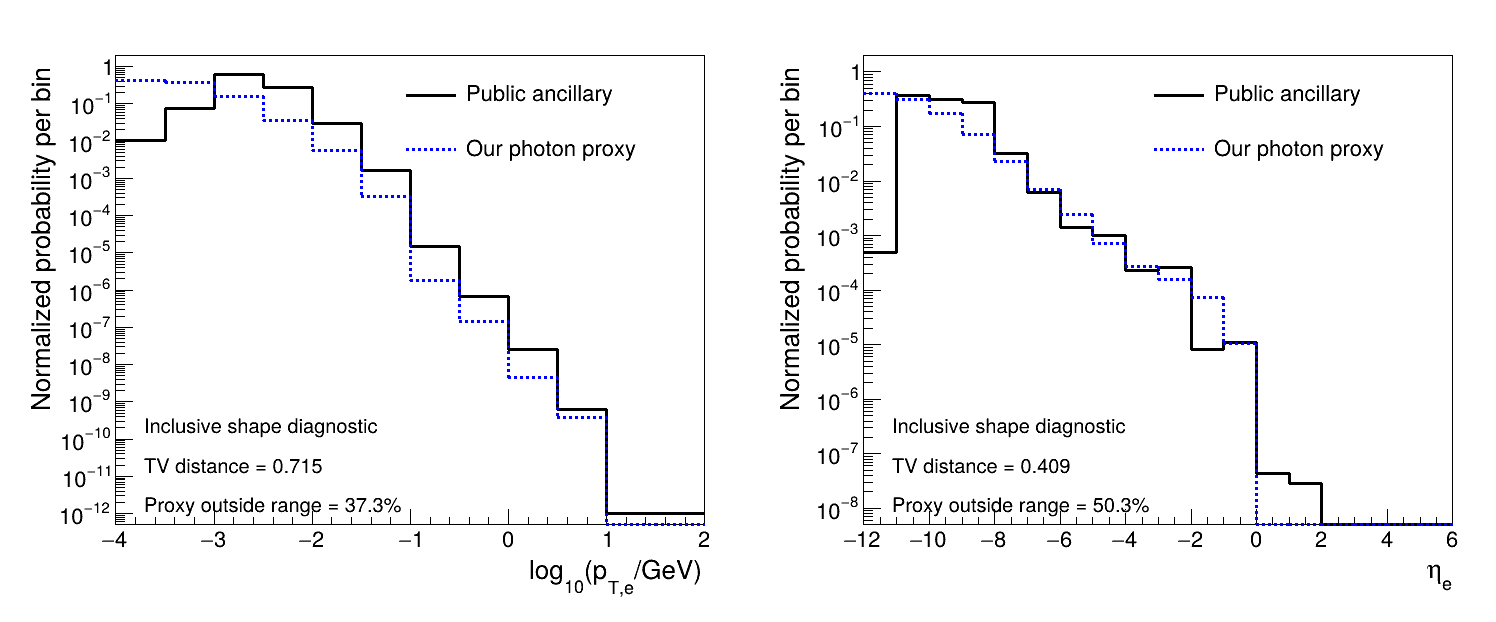}
\caption{Inclusive recoil-electron distributions for the public ancillary photon sample (black) and our photon proxy (blue), normalized separately within the displayed ranges. Annotations give the total-variation distances and the fractions of proxy weight outside those ranges.}\label{fig:3}
\end{figure*}

Figure~\ref{fig:3} compares our generated photon distributions with the public ancillary distributions under the same beam kinematics. The distributions exhibit similar overall trends, with quantitative differences after normalization within the displayed ranges. Our integrated photon rates exceed the published estimates by factors of 3.6--4.1; for the classifier comparison, we normalize the photon contribution to the published effective rate after preselection. The results remain conditional on our generated background model.

The public distributions are not used to train either selector or to reweight our generated kinematic distributions.

\subsection{DIS contribution}\label{sec:3.3}

A separate rate-level cross-check was completed with MadGraph5\_aMC@NLO 3.7.0 for leading-order $ep\to ej$ using CT10nlo, scaled by A = 197, 5\% veto survival, and the stated $10^{-6}$ jet-miss factor. Across the 11 Table I mass-dependent selections, our effective DIS estimates differ from the published values by 4.30\% to 6.23\%.

No event-level DIS sample is classified in this study. We retain the published effective DIS rate after preselection and set its second-stage selection efficiency to unity for both methods, so neither selector receives credit for additional DIS rejection. Including this unrejected component reduces the improvement relative to the unrefined preselection compared with a photon-only calculation. However, it does not guarantee a conservative BDT-versus-cuts comparison: the actual DIS efficiencies could differ between methods and alter their relative performance.

\section{Fair Machine-Learning Test}\label{sec:4}

\subsection{Two nested information sets}\label{sec:4.1}

The primary, electron-only input set contains $\log_{10}$($Q^2$), $p_{T,e}$, $\eta_e$, and $E_e$. These quantities correspond to measurements that can in principle be reconstructed from the outgoing electron. In the present study they are evaluated exactly from generator-level four-momenta: no tracking efficiency, energy scale, angular resolution, bremsstrahlung recovery, acceptance loss, or variable-to-variable detector correlation is applied. At fixed beam energy, $p_{T,e}$ and $\eta_e$ already determine most of the outgoing-electron four-momentum, so $E_e$ and $Q^2$ are substantially redundant. This makes the electron-only test intentionally demanding for ML: the model receives several representations of roughly two independent electron degrees of freedom.

The extended input set contains the same four electron quantities plus $t$. Both the BDT and rectangular selector receive the same inputs in each comparison, so the change in their relative performance measures how they use $t$.

\subsection{Identical data roles and prevention of test leakage}\label{sec:4.2}

For every signal type, boson mass, and input set, the BDT and rectangular selector use the same weighted signal and photon events. Independently generated samples have distinct roles. The training sample is used to fit the BDT and construct rectangular candidates. The validation sample selects the BDT profile, BDT score threshold, rectangular grid, and final box boundaries. These choices are finalized using the validation sample before evaluation on the independent test sample and are not adjusted using the test results.

The held-out test events are used once for the reported comparison. The two methods are evaluated on the same test events, so their difference can be estimated with a paired bootstrap instead of treating the two results as statistically independent. Source files, executables, configurations, selected models, cut boxes, generated data, and final results are recorded in checksum manifests. This design limits accidental leakage, post-test retuning, and comparisons between mismatched generator versions.

\subsection{BDT and rectangular-selector definitions}\label{sec:4.3}

The ML method is a boosted decision tree implemented with TMVA. At each of the 11 Table I masses for both vector and scalar signals, four prespecified model profiles are considered: 400-tree models with maximum depths two, three, and four, and an 800-tree depth-three model. Training fits the trees, and validation selects among these fixed profiles and chooses the score threshold that maximizes the common analysis objective.

The classical comparator is not merely the unchanged Table I selection. It is a weighted multistart rectangular selector. Candidate lower and upper bounds are generated from weighted-threshold grids containing 40, 80, and 160 quantiles, with multiple starting points to reduce dependence on a single greedy solution. Training constructs candidate boxes and validation selects the final grid and bounds. The method is a strong approximate rectangular baseline, although it is not claimed to be the mathematically global optimum over every possible box.

\subsection{Optimization target, uncertainty, and decision rules}\label{sec:4.4}

Both methods act only as second-stage refinements after the published Table I preselection. They maximize the same approximate background-dominated significance ratio

\begin{equation} R_Z = \epsilon_S\sqrt{\frac{\sigma_\gamma+\sigma_{\mathrm{DIS}}}{\epsilon_\gamma\sigma_\gamma+\sigma_{\mathrm{DIS}}}}.\label{eq:rz}\end{equation}

Here $\epsilon_S$ and $\epsilon_\gamma$ are the weighted signal and photon efficiencies of the second-stage selector, while $\sigma_\gamma$ and $\sigma_{\mathrm{DIS}}$ are the effective Table I background cross sections. The DIS term is left unrejected. $R_Z$ = 1 means that the second-stage selector provides no improvement relative to stopping after Table I; values above one indicate improvement within this proxy. The direct algorithm comparison is the ratio $R_Z$(BDT)/$R_Z$(rectangular).

\begin{equation} A_g=1-\sqrt{\frac{R_Z(\mathrm{rectangular})}{R_Z(\mathrm{BDT})}}.\label{eq:ag}\end{equation}

Because the signal cross section scales as the square of the electron-boson coupling in this model, $A_g$ is the relative reduction in the coupling threshold implied by the ratio of the two approximate significance metrics. Positive values favor the BDT and negative values favor rectangular cuts. This quantity is an algorithmic coupling-threshold proxy, not an absolute exclusion or discovery reach.

A +1\% coupling advantage was fixed before the final robustness evaluation as the minimum practically useful effect. This is a project decision threshold, not a universal EIC or community standard. Through the equation above, a 1\% coupling advantage corresponds to approximately a 2.03\% increase in $R_Z$ over the rectangular selector. The threshold prevents a statistically resolvable but scientifically negligible sub-percent fluctuation from being described as a useful ML improvement.

The full electron-only mass scan contains 22 primary comparisons: 11 masses for each of two signal spins. Paired event bootstraps provide pointwise intervals, and a Bonferroni lower bound controls the family of primary tests. A separate robustness study repeats the complete event generation, training, validation, model selection, cut optimization, and sealed-test evaluation ten times at four prespecified electron-only benchmarks: vector masses 1 and 10 GeV and scalar masses 1 and 6.31 GeV. It contains 40 complete replicas, 2,000 paired event-bootstrap samples per replica, and 20,000 hierarchical bootstrap draws per benchmark.

The $t$-inclusive robustness study repeats the same protocol at the two positive high-mass benchmarks: vector 10 GeV and scalar 10 GeV. It contains ten complete replicas per signal, 2,000 paired event-bootstrap samples per replica, and 20,000 hierarchical bootstrap draws per signal. Every replica regenerates independent training, validation, and test events, refits all four BDT profiles, and reoptimizes all three rectangular grid settings. All method and result checksum manifests pass verification.

\section{Results}\label{sec:5}

\subsection{Weighted kinematic structure after Table I}\label{sec:5.1}

\begin{figure*}[!htbp]
\centering
\includegraphics[width=\textwidth,height=0.42\textheight,keepaspectratio]{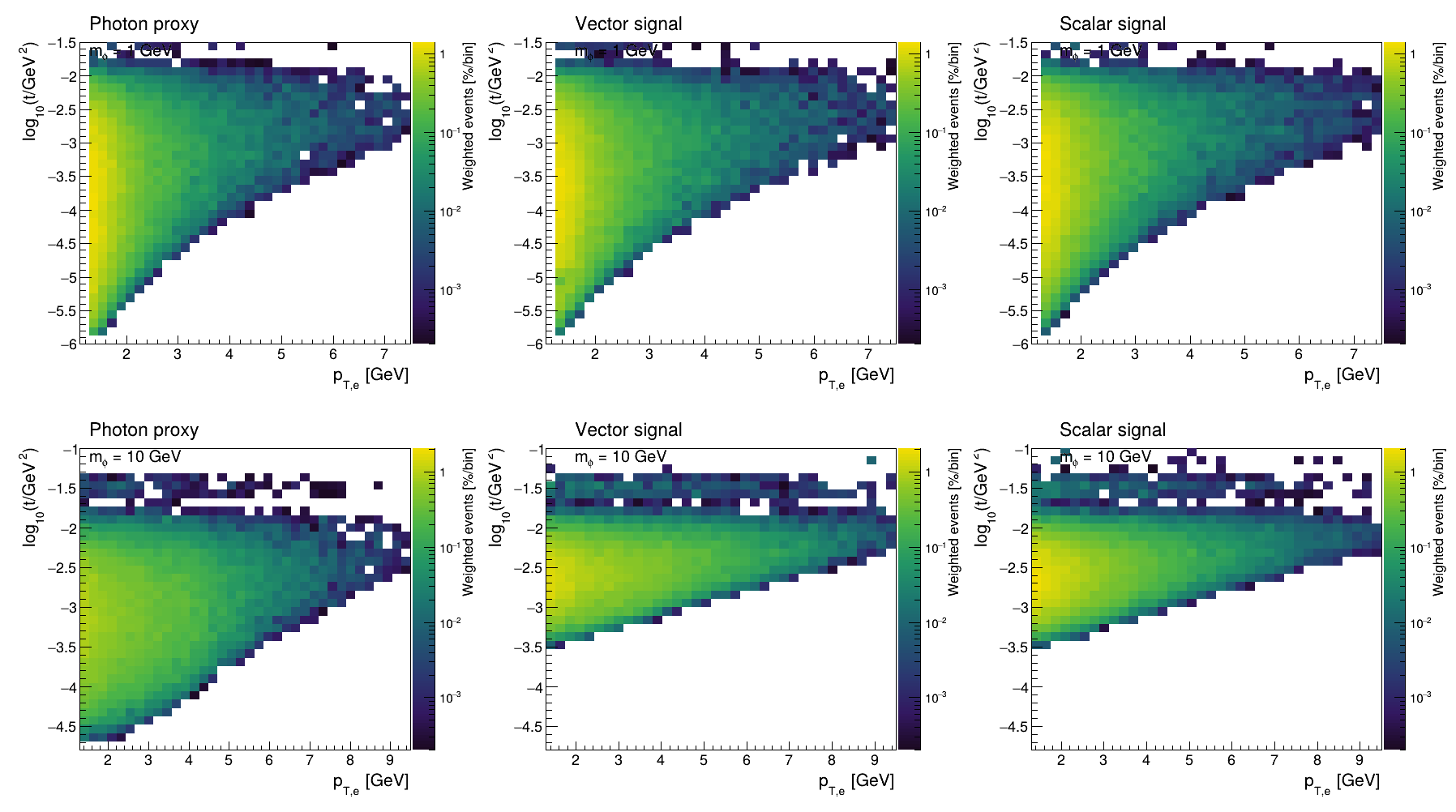}
\caption{Weighted post-Table-I distributions in $\log_{10}$($t$/$\mathrm{GeV}^2$) and $p_{T,e}$. Rows correspond to $m_\phi$ = 1 and 10 GeV; columns show the coherent-photon proxy, vector signal, and scalar signal. Each panel is normalized to its total weighted class probability, so color represents shape rather than absolute cross section. Signal events use their integration weights, while photon events additionally include the stated photon-miss probability. Identical axes, bins, and color scales are used within each mass row.}\label{fig:4}
\end{figure*}

Figure~\ref{fig:4} shows the weighted joint distributions of nuclear momentum transfer and recoil-electron transverse momentum for the photon proxy and both signal hypotheses at 1 and 10 GeV.

At 1 GeV, all three classes populate broadly similar wedge-shaped regions in the $t$-$p_{T,e}$ plane. At 10 GeV, the photon proxy retains substantially more weight at smaller $t$, while both signal hypotheses shift toward larger $t$ and exhibit a different correlation with $p_{T,e}$. A single independent interval in each variable cannot follow the complete sloped boundary. This visual pattern is consistent with, but does not by itself prove, the later finding that $t$ is useful to the nonlinear BDT mainly at high mass.

\subsection{Impact of \texorpdfstring{$t$}{t} across boson mass}\label{sec:5.2}

\begin{figure*}[!htbp]
\centering
\includegraphics[width=\textwidth,height=0.42\textheight,keepaspectratio]{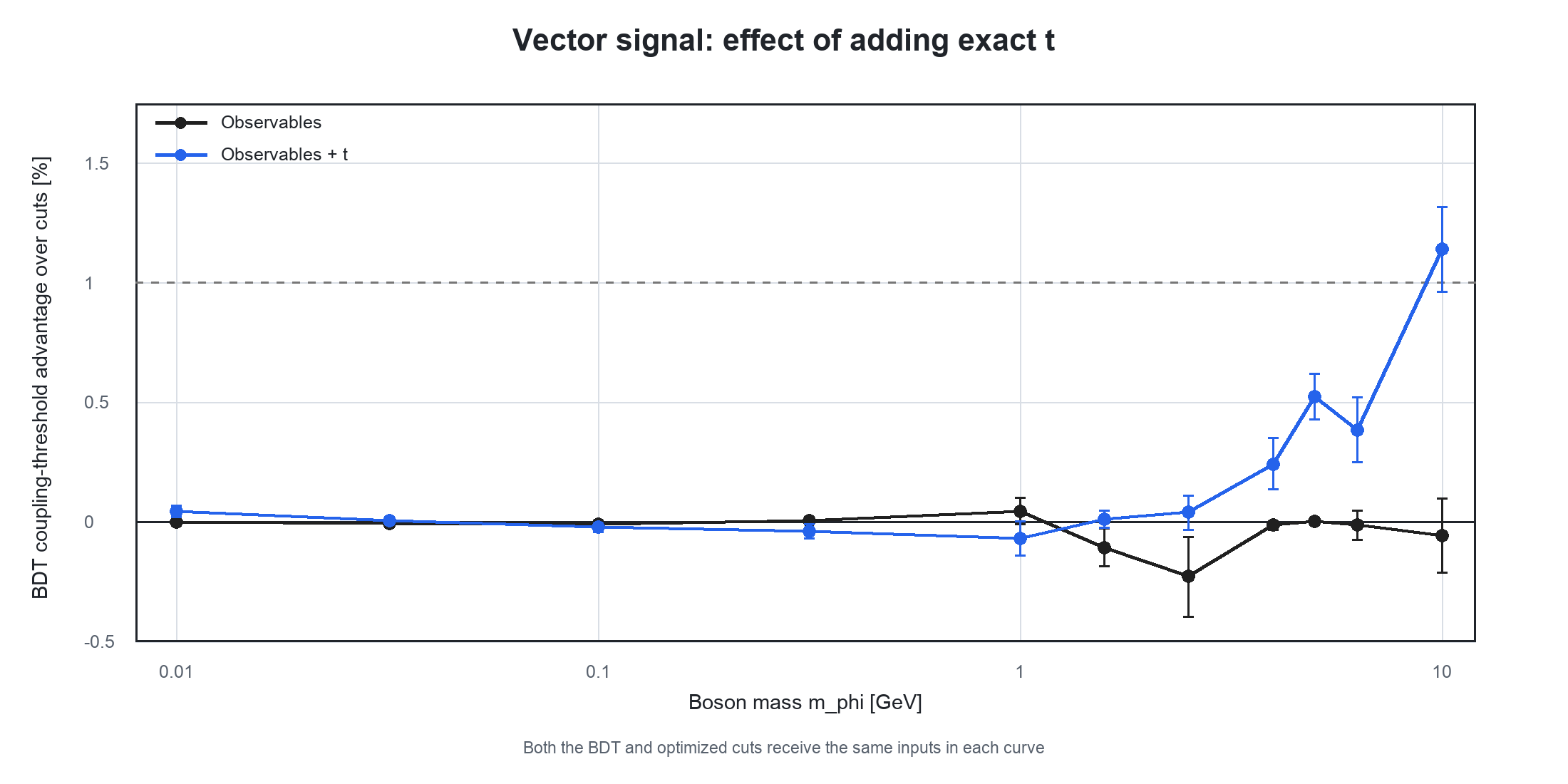}
\caption{Vector mass scan. The logarithmic horizontal axis is the vector-boson mass. The vertical axis is the BDT coupling-threshold advantage over same-input optimized rectangular cuts; zero denotes equal performance and positive values favor the BDT. Solid black circles use $Q^2$, $p_{T,e}$, $\eta_e$, and $E_e$. Solid blue circles give both algorithms the same four variables plus $t$. Error bars are paired 95\% event-bootstrap intervals. The dotted horizontal line marks the predefined +1\% practical threshold.}\label{fig:5}
\end{figure*}

\begin{figure*}[!htbp]
\centering
\includegraphics[width=\textwidth,height=0.42\textheight,keepaspectratio]{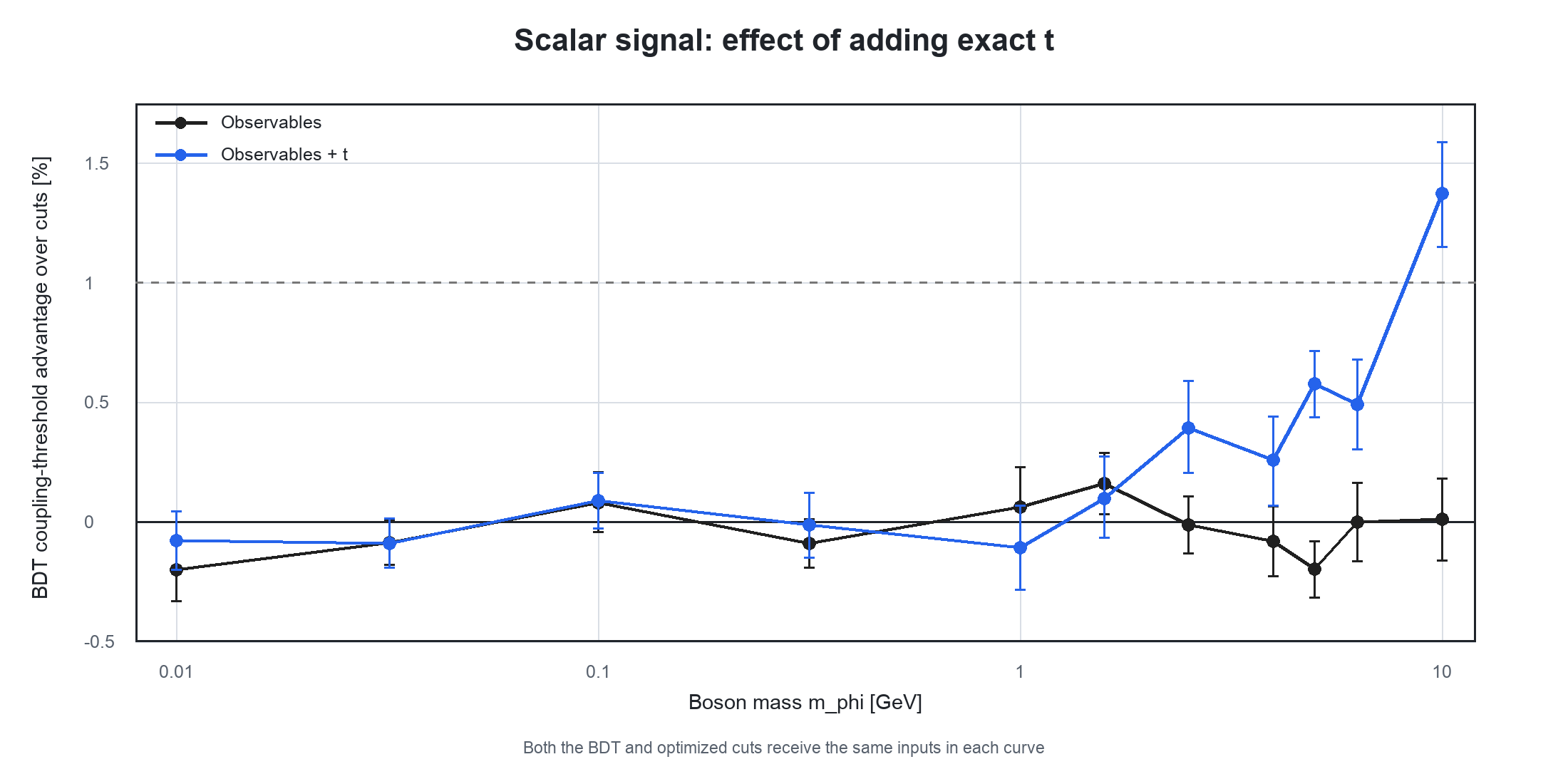}
\caption{Scalar mass scan. Axes and statistical conventions match Figure~\ref{fig:5}. Solid black circles use the four recoil-electron inputs. Solid blue circles add $t$ to both the BDT and optimized rectangular selector.}\label{fig:6}
\end{figure*}

Figures~\ref{fig:5} and~\ref{fig:6} show the mass dependence of the BDT coupling-threshold advantage $A_g$ for vector and scalar signals. The paired 95\% event-bootstrap intervals describe uncertainty for each selected model and cut box evaluated on the independent test sample; variation from new event generation and retraining is examined in Section~\ref{sec:5.4}.

For the vector signal, the electron-only points remain close to zero at every mass. With $t$, the advantage is also small at low mass but increases toward the upper end of the scan. This pattern suggests that  $t$ becomes more useful to the BDT at high mass.

The scalar scan shows the same qualitative behavior: electron-only differences fluctuate around zero, while adding $t$ produces a clearer positive trend above approximately 2 GeV.

Across all 22 electron-only scalar/vector points, the nominal coupling-threshold difference ranges from -0.228\% to +0.161\%. No point has a positive familywise-controlled lower bound, and no point reaches the +1\% practical threshold. The largest nominal positive electron-only result occurs for the 1.585 GeV scalar: +0.161\%, with a pointwise 95\% interval from +0.033\% to +0.289\%, but its familywise lower bound is -0.026\%. Several points slightly favor rectangular cuts. The complete scan therefore provides no multiplicity-controlled evidence that an electron-only BDT is better.

This conclusion does not mean that the Table I selection is already optimal. Both second-stage methods can improve the scalar proxy relative to leaving Table I unchanged. Around 1--2.5 GeV, their $R_Z$ values reach approximately 1.05--1.11. The important result is that the optimized rectangular selector captures essentially all of this electron-only improvement. The gain is due to reoptimizing the selection on our samples, not specifically due to machine learning.

The $t$-inclusive curves answer a different question. In the single sealed mass scan at 10 GeV, the BDT advantage is +1.14\% with a pointwise 95\% interval of [+0.96\%, +1.32\%] for the vector and +1.37\% with an interval of [+1.15\%, +1.59\%] for the scalar. These single-sample intervals alone would not demonstrate that the effect survives new event generation and retraining. Section~\ref{sec:5.4} therefore repeats the entire analysis across independent seeds.

\subsection{Operating characteristics and variable attribution at 10 GeV}\label{sec:5.3}

\begin{figure*}[!htbp]
\centering
\includegraphics[width=\textwidth,height=0.42\textheight,keepaspectratio]{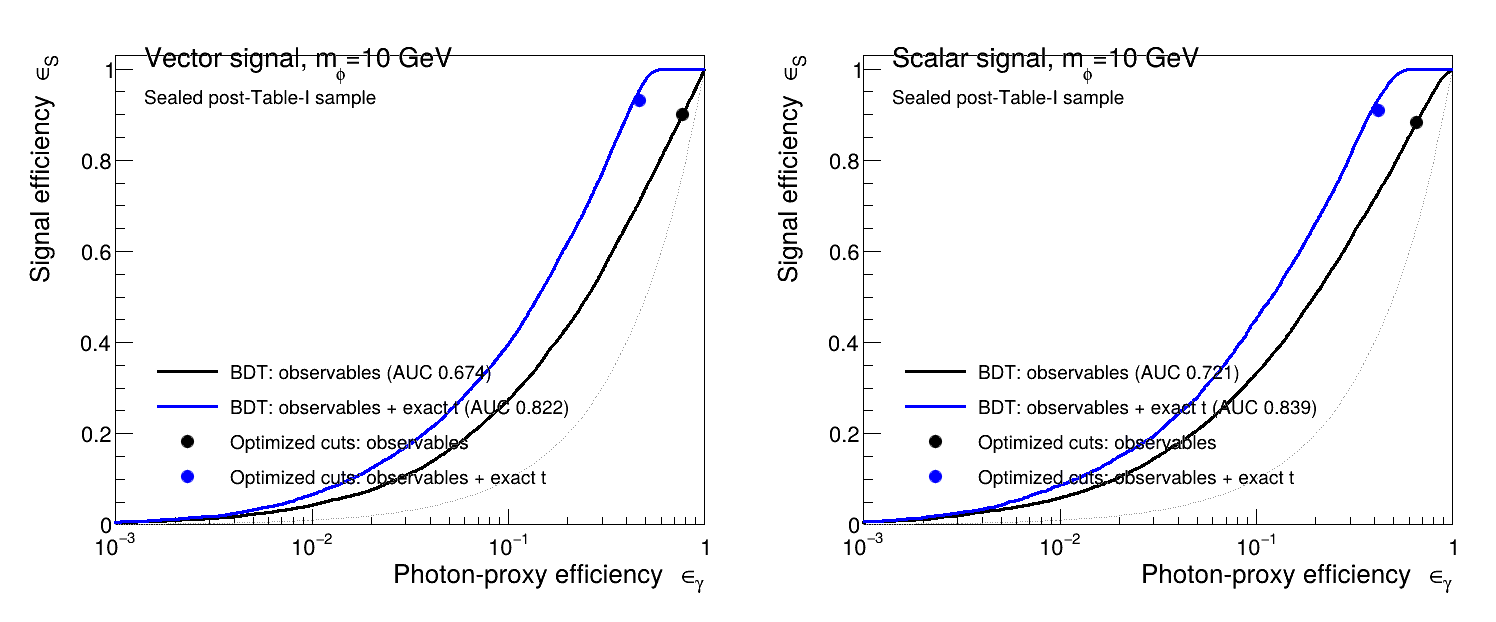}
\caption{Weighted post-Table-I signal efficiency $\epsilon_S$ versus coherent-photon-proxy efficiency $\epsilon_\gamma$ on the sealed 10 GeV test samples. The left and right panels show vector and scalar signals. Solid curves vary the score threshold of the unchanged trained BDTs; filled circles show the optimized rectangular operating points chosen using the validation sample. Black denotes $Q^2$, $p_{T,e}$, $\eta_e$, and $E_e$, while blue adds $t$ to both methods. DIS is absent from this diagnostic because no event-level DIS sample is classified.}\label{fig:7}
\end{figure*}

Figure~\ref{fig:7} separates the two ingredients of the final comparison. A continuous BDT curve is obtained by varying the score threshold on the independent post-Table-I test events while keeping the trained model unchanged. This scan describes performance across thresholds; the threshold used for the reported selection is chosen using the validation sample. Each optimized rectangular selector supplies one validation-selected operating point rather than a continuous ROC curve. The horizontal axis is the weighted survival efficiency of the coherent-photon proxy, $\epsilon_\gamma$, and the vertical axis is weighted signal efficiency, $\epsilon_S$. Curves closer to the upper-left corner rank signal above photon more effectively. The plot contains no event-level DIS because no such sample is classified in this study.

For the vector signal, $t$ raises the sealed weighted AUC from 0.674 to 0.822; for the scalar it raises the AUC from 0.721 to 0.839. The blue curves therefore show a substantial improvement in signal-photon ranking across thresholds, even though the final coupling-threshold advantage over a same-input optimized box is only 1.1\% to 1.4\%. AUC and coupling advantage answer different questions: AUC integrates ranking performance over every threshold, whereas $A_g$ compares two validation-selected working points after the unrejected DIS rate is included in $R_Z$.

\begin{figure*}[!htbp]
\centering
\includegraphics[width=\textwidth,height=0.42\textheight,keepaspectratio]{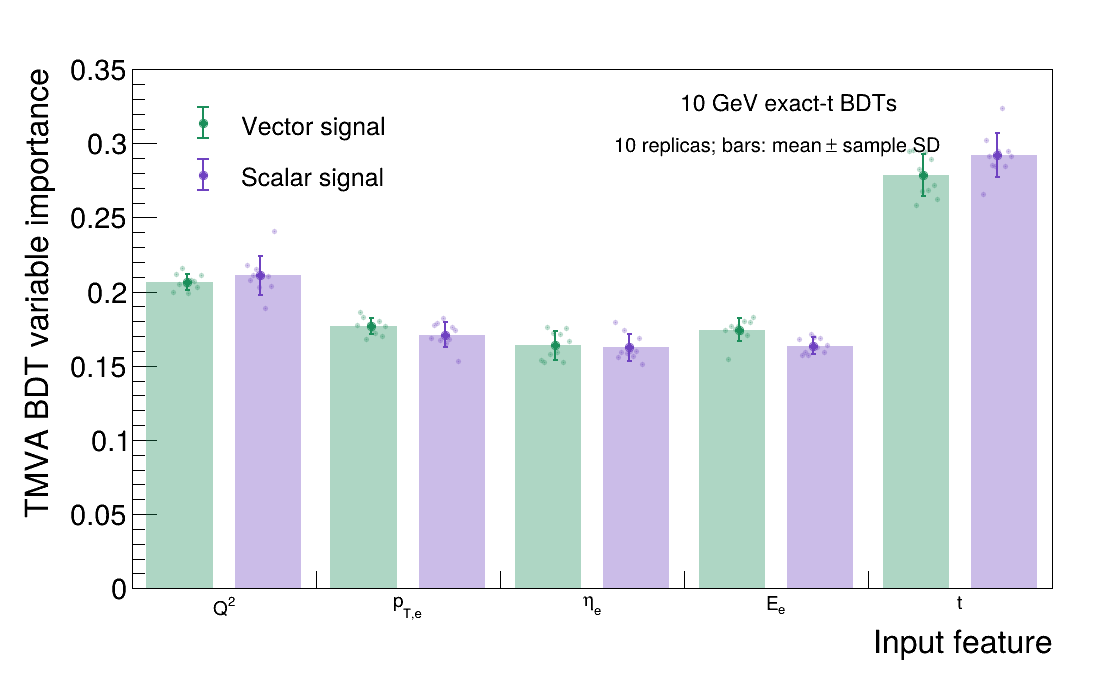}
\caption{TMVA BDT variable importance for the validation-selected $t$-inclusive models across ten independent 10 GeV generation-and-training replicas. Green and purple denote vector and scalar signals. Small points show individual replicas; large circles and error bars show the mean and sample standard deviation. Importance is normalized within each fitted model. }\label{fig:8}
\end{figure*}

Figure~\ref{fig:8} shows the normalized TMVA variable importances for the models selected using validation data in the ten independent 10 GeV replicas.

$t$ is ranked highest in both ensembles. Its mean importance is 0.2786 $\pm$ 0.0143 for the vector and 0.2920 $\pm$ 0.0148 for the scalar. The remaining mean importances span 0.164 to 0.212. This ranking is consistent with the density and ROC plots: the positive high-mass result is associated with $t$ rather than with a different electron-only model choice. However, TMVA importance is training-derived and correlation-dependent. It neither proves causation nor shows that a realistically reconstructed $t$ would retain the same value.

\subsection{Independent-seed robustness}\label{sec:5.4}

To assess dependence on the generated samples and training seeds, we repeat the complete generation, training, model selection, cut optimization, and independent test evaluation using the protocol in Section~\ref{sec:4.4}. This tests variation beyond resampling one fixed test sample.

\begin{figure*}[!htbp]
\centering
\includegraphics[width=\textwidth,height=0.42\textheight,keepaspectratio]{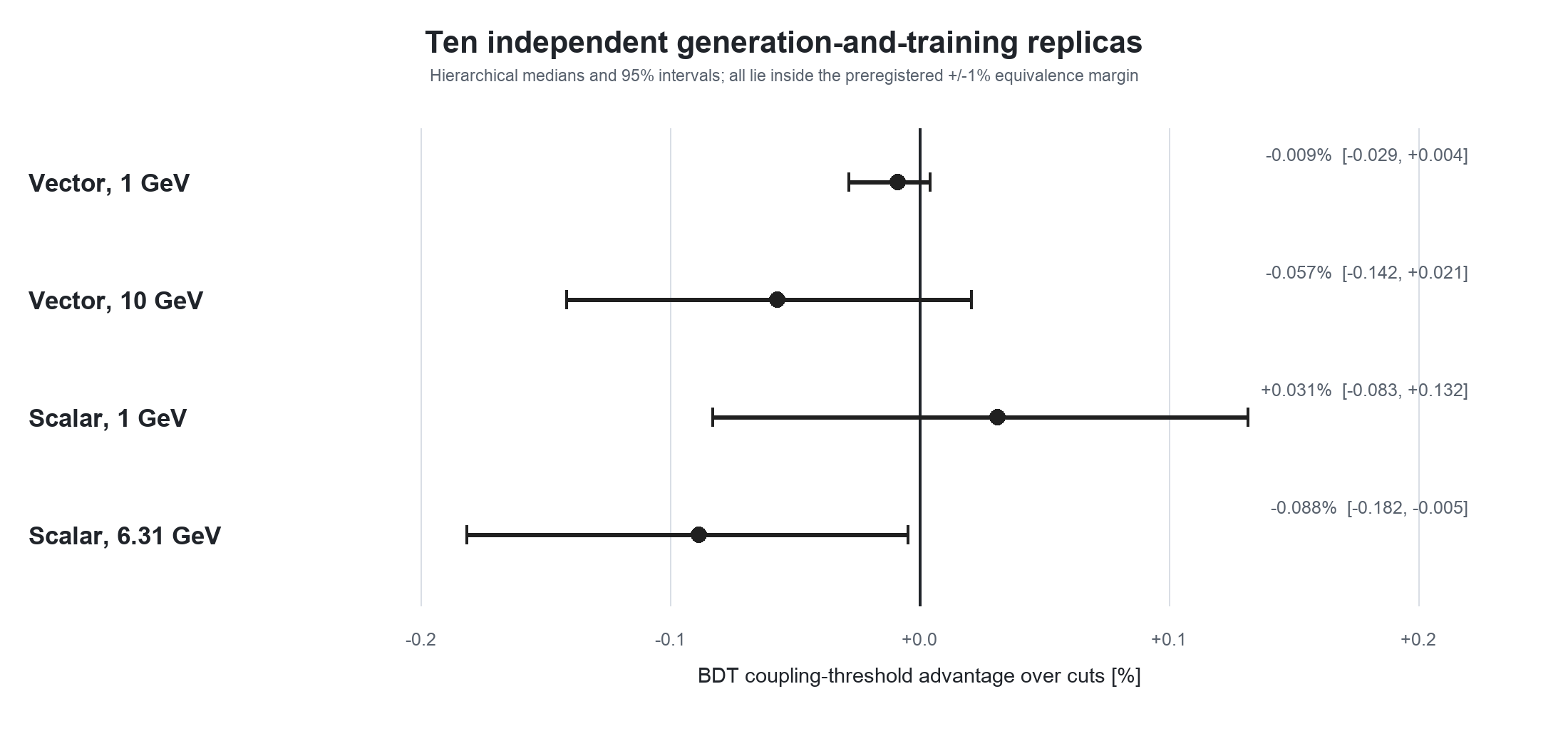}
\caption{Ten-replica electron-only robustness comparison. Each row is one prespecified signal benchmark. The horizontal axis is the BDT coupling-threshold advantage over same-input optimized rectangular cuts; zero means equal performance. Each point is the hierarchical median over ten complete generation-and-training replicas, and each horizontal bar is the corresponding 95\% interval. All four intervals lie far inside the predefined -1\% to +1\% practical-equivalence region.}\label{fig:9}
\end{figure*}

\begin{table*}[!htbp]
\caption{Independent-seed results. All four ordinary 95\% intervals lie inside the preregistered -1\% to +1\% practical-equivalence range; none passes the familywise superiority test.}\label{tab:2}
\centering\small\setlength{\tabcolsep}{4pt}
\begin{tabular}{lcccc}
\toprule
Signal & $m_\phi$ [GeV] & BDT coupling advantage [95\%] & BDT wins & Decision \\
\midrule
Vector & 1.000 & -0.009\% [-0.029\%, +0.004\%] & 4/10 & Practical equivalence \\
Vector & 10.000 & -0.057\% [-0.142\%, +0.021\%] & 2/10 & Practical equivalence \\
Scalar & 1.000 & +0.031\% [-0.083\%, +0.132\%] & 8/10 & Practical equivalence \\
Scalar & 6.310 & -0.088\% [-0.182\%, -0.005\%] & 0/10 & Cuts slightly favored \\
\bottomrule\end{tabular}
\end{table*}

Figure~\ref{fig:9} and Table~\ref{tab:2} show that the electron-only conclusion persists across independent generation-and-training replicas. Vector 1 GeV, vector 10 GeV, scalar 1 GeV, and scalar 6.31 GeV all fall within the preregistered practical-equivalence margin. No benchmark has a familywise lower bound above zero, and none approaches a +1\% BDT advantage. Scalar 6.31 GeV has a 95\% interval slightly below zero, so optimized cuts are favored statistically in that isolated comparison, but the magnitude is only about one tenth of one percent and remains scientifically negligible under the predefined margin.

Selected BDT profiles and rectangular grid sizes vary among replicas because their validation scores are nearly degenerate, yet the final performance ratio remains near one. Practical equivalence therefore persists across these changes in the selected configurations.

\begin{figure*}[!htbp]
\centering
\includegraphics[width=\textwidth,height=0.42\textheight,keepaspectratio]{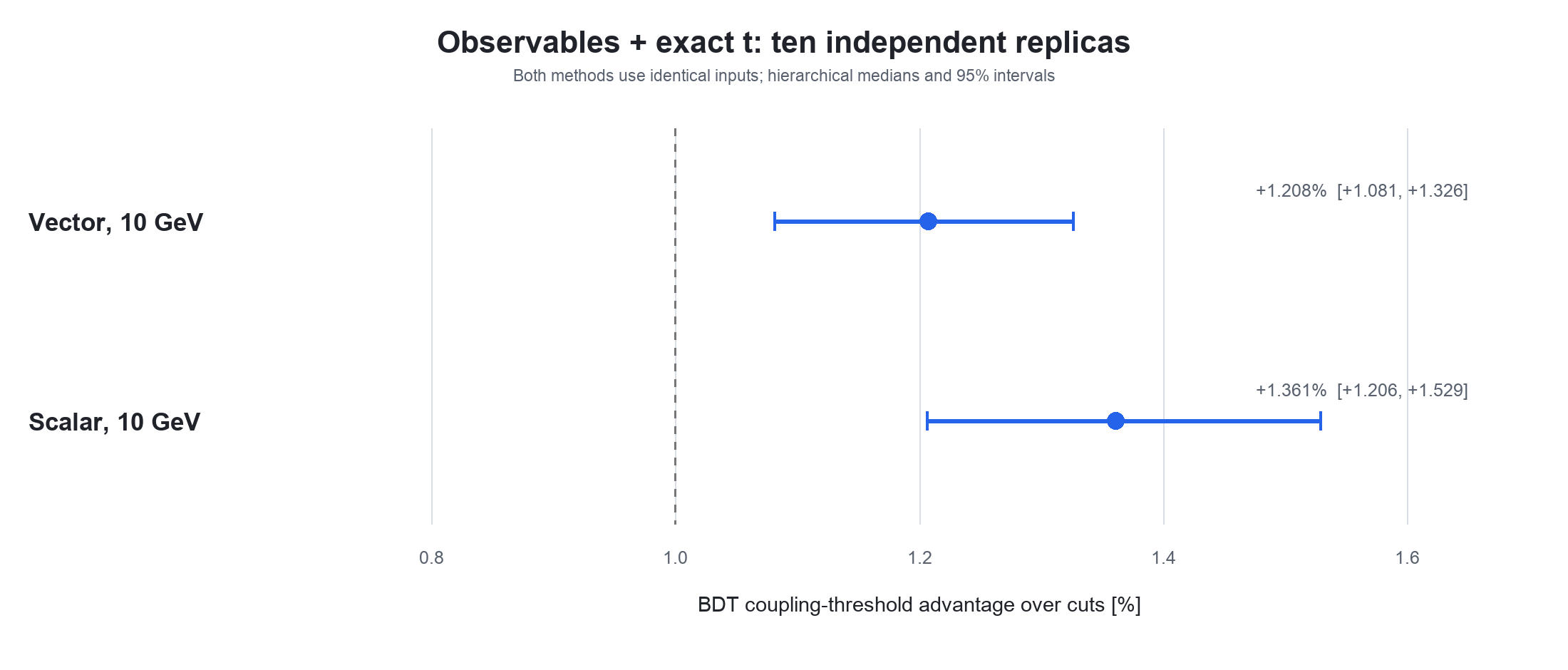}
\caption{Ten-replica $t$-inclusive robustness comparison at 10 GeV. The horizontal axis is the BDT coupling-threshold advantage over same-input optimized cuts. Blue circles are hierarchical medians and horizontal bars are 95\% intervals over complete independent generation-and-training replicas. The dotted vertical line marks the predefined +1\% practical threshold. Both methods receive $Q^2$, $p_{T,e}$, $\eta_e$, $E_e$, and $t$.}\label{fig:10}
\end{figure*}

\begin{table*}[!htbp]
\caption{Independent-seed $t$-inclusive results. Hierarchical paired intervals combine complete generator-and-training replicas with paired event resampling. Both familywise lower bounds exceed the predefined +1\% practical threshold.}\label{tab:3}
\centering\small\setlength{\tabcolsep}{4pt}
\begin{tabular}{lccccc}
\toprule
Signal & $m_\phi$ [GeV] & BDT advantage [95\%] & BDT wins & Familywise lower & Decision \\
\midrule
Vector & 10.000 & +1.208\% [+1.081\%, +1.326\%] & 10/10 & +1.081\% & Practical BDT superiority \\
Scalar & 10.000 & +1.361\% [+1.206\%, +1.529\%] & 10/10 & +1.206\% & Practical BDT superiority \\
\bottomrule\end{tabular}
\end{table*}

Figure~\ref{fig:10} and Table~\ref{tab:3} show that the $t$-inclusive result survives every independent replica. The BDT wins 10 of 10 vector comparisons and 10 of 10 scalar comparisons. The hierarchical median coupling-threshold advantages are +1.208\% [1.081\%, 1.326\%] for the vector and +1.361\% [1.206\%, 1.529\%] for the scalar. Their two-test familywise lower bounds are +1.081\% and +1.206\%, respectively. Both lower bounds remain above the predefined +1\% threshold. This passes the statistical-superiority and practical-usefulness decision rules.

Together, Figures~\ref{fig:9} and~\ref{fig:10} show practical equivalence with electron-only inputs and a modest, consistent BDT advantage when both selectors also receive $t$.

\subsection{Size and scientific meaning of the high-mass effect}\label{sec:5.5}

The median coupling-threshold advantages of +1.208\% for the vector and +1.361\% for the scalar correspond to $R_Z(\mathrm{BDT})/R_Z(\mathrm{rectangular})$ values of approximately 1.0246 and 1.0278, respectively. Thus, the $t$-inclusive BDT improves the significance proxy by roughly 2.5\% and 2.8\% relative to optimized cuts at 10 GeV. The hierarchical intervals account for finite simulated-event variation and generation/training-seed variation. They do not include detector or physics-model systematic uncertainties, discussed in Section~\ref{sec:6.2}. The result establishes a small, repeatable advantage and motivates testing its survival with reconstructed $t$.

\section{Discussion and Next Stage}\label{sec:6}

\subsection{Why \texorpdfstring{$t$}{t} changes the machine-learning comparison}\label{sec:6.1}

The two input sets give an information-based explanation for the otherwise different conclusions. The electron-only variables all describe one recoil electron. At fixed incoming energy, $p_{T,e}$ and $\eta_e$ determine the electron direction and most of its energy-momentum, while $E_e$ and $Q^2$ provide correlated representations of the same object. The optimized rectangular selector achieves nearly the same performance as the BDT with electron-only inputs. A BDT can represent more complicated boundaries, but it cannot create information that is absent from its inputs.

$t$ adds a variable from the intact-ion side of the exclusive event. Because $\phi$ is invisible, the electron alone does not completely determine the recoil configuration. The relationship among $t$ and the electron quantities changes across signal and coherent-photon phase space, especially toward high boson mass. A BDT can partition this multidimensional space into several correlated regions, whereas a rectangular selector can only impose independent upper and lower bounds on each variable. The 10 GeV results indicate that this nonlinear combination leaves a small residual advantage after the box has been strongly optimized.

Detector acceptance and resolution may change the correlations responsible for the advantage. The next question is whether reconstructed $t$ preserves enough of this structure for the BDT advantage to survive.

\subsection{Physics limitations}\label{sec:6.2}

The scalar matrix element has not been checked against an independent analytic calculation or generator implementation. The background model remains a principal source of uncertainty. As discussed in Section~\ref{sec:3.2}, the generated and public photon distributions have similar overall trends but differ quantitatively in shape and rate. Normalizing to the published effective cross section fixes the rate after preselection, while the multivariate correlations used by the selectors remain those of our generator. 

The DIS rate cross-check is described in Section~\ref{sec:3.3}, but no event-level DIS sample is classified. Leaving the published DIS rate unrejected is conservative relative to the unrefined Table I result for either selector separately, but it does not prove that the real BDT-versus-cuts ordering is conservative because the two unknown DIS efficiencies may differ. Rare photon-miss probabilities near $10^{-6}$ also cannot be validated by ordinary samples containing only thousands or millions of events without a detector-tail model or control-region strategy.

No detector response is applied. The analysis contains no far-forward ion acceptance, beam optics, reconstruction efficiency, $t$ smearing, photon-veto response, zero-degree-calorimeter resolution, pileup, beam background, or detector-induced correlation.

Finally, $R_Z$ is a background-dominated $S/\sqrt{B}$-type proxy rather than a nuisance-aware profile likelihood. The paired and hierarchical intervals quantify finite simulated-event and observed generation/training-seed variation. They do not include detector, generator, luminosity, nuclear-model, or background-normalization systematics.

\subsection{Detector simulation with reconstructed \texorpdfstring{$t$}{t}}\label{sec:6.3}

The next step is to repeat the comparison with detector response included. A detector simulation would model the measurement of the recoil electron and outgoing ion, including acceptance, reconstruction efficiency, and momentum resolution, and determine how accurately $t$ can be reconstructed. The study should report the acceptance, efficiency, bias, and resolution of reconstructed $t$ as functions of boson mass and true $t$, and account for photon-veto and zero-degree-calorimeter responses.

Both the BDT and optimized rectangular selector would use the same reconstructed inputs, following the training, validation, independent-test, and replica procedures described in Section~\ref{sec:4}. Comparing their performance with the present results would establish whether the observed advantage survives realistic measurement effects. The study would also assess background-model uncertainties and systematic effects.

\section{Conclusions}\label{sec:7}

We performed a controlled comparison of boosted decision trees and optimized rectangular cuts for coherent exclusive scalar and vector dark-boson production at the EIC. With electron-only inputs, the BDT provides no practically useful improvement over optimized rectangular cuts. The 22-point mass scan finds no familywise-controlled evidence of BDT superiority, and ten independent generation-and-training replicas at each of four benchmarks place all ordinary 95\% intervals within the predefined $-1\%$ to $+1\%$ practical-equivalence range.

Adding $t$ to both methods produces median coupling-threshold advantages at 10 GeV of +1.208\% for the vector and +1.361\% for the scalar, with hierarchical 95\% intervals of [1.081\%, 1.326\%] and [1.206\%, 1.529\%], respectively. The BDT wins all ten independently generated and retrained replicas for both signals, and both familywise lower bounds exceed the predefined +1\% practical threshold.

These results show that adding $t$ creates a modest, repeatable BDT advantage. The next step is a detector simulation comparing BDTs and reoptimized rectangular cuts using reconstructed electron variables and $t$, to determine whether the advantage survives realistic acceptance and measurement resolution.

\section*{Author Contributions}
Rojae Mighty: Conceptualization, methodology, software, validation, formal analysis, visualization, data curation, and writing---original draft. Ankush Reddy Kanuganti: Supervision, conceptualization, and writing---review and editing.

\section*{Data Availability}
Github repository with code, data, and results: \url{https://github.com/iRojae/Dark-Boson-ML-EIC}.

\section*{Acknowledgments}
The author thanks Hongkai Liu for helpful discussions of coherent dark-boson production. This work was supported by the Office of Nuclear Physics within the U.S. Department of Energy’s
Office of Science.

\section*{Conflicts of Interest}
Rojae Mighty declares no conflict of interest. Ankush Reddy Kanuganti declares no conflict of interest.

\end{document}